\documentclass[11pt, a4paper]{article}
\pdfoutput=1
\usepackage{jcappub}

\usepackage{booktabs}
\usepackage{multirow}
\usepackage{amssymb}
\usepackage[export]{adjustbox}
\usepackage{floatrow}
\newfloatcommand{capbtabbox}{table}[][3.5in]
\newfloatcommand{capbfigbox}{figure}[][2.3in]
\usepackage{blindtext}
\usepackage{amsmath}
\usepackage{booktabs}
\usepackage{aas_macros}

\newcommand{\ie}{i.e.~}

\def\lsim{\mathrel{\raise.3ex\hbox{$<$\kern-.75em\lower1ex\hbox{$\sim$}}}}
\def\gsim{\mathrel{\raise.3ex\hbox{$>$\kern-.75em\lower1ex\hbox{$\sim$}}}}

\usepackage{appendix}
\usepackage{array}
\newcolumntype{x}[1]{>{\centering\arraybackslash}p{#1}}
\usepackage{bm}
\usepackage{color}
\usepackage{graphicx}
\usepackage{cancel}
\usepackage{amsfonts}
\usepackage{amssymb}
\usepackage{amsmath}
\usepackage{calc}
\usepackage{dcolumn}
\usepackage{mathrsfs}
\usepackage{mathtools}
\usepackage{soul}

\usepackage[normalem]{ulem}

\newcommand{\Ref}[1]{Ref.~\cite{#1}}     

\newcommand{\beq}{\begin{equation}}
\newcommand{\eeq}{\end{equation}}

\definecolor{rossoCP3}{cmyk}{0,.88,.77,.40}
\definecolor{verdeCP3}{rgb}{0.09765625, 0.57421875, 0.1015625}
\definecolor{bluCP3}{rgb}{0, 0.23, 0.67}

\ifx\pdfoutput\undefined
\usepackage[dvips,bookmarks]{hyperref}    
\else
\usepackage{hyperref}    
\fi
\hypersetup{colorlinks, bookmarksopen, bookmarksnumbered,
citecolor=verdeCP3, linkcolor=bluCP3, pdfstartview=FitH, urlcolor=rossoCP3}

\begin{document}

\hspace*{110mm}{\large \tt FERMILAB-PUB-17-390-A}

\vskip 0.2in

\title{Resolving Dark Matter Subhalos \\ With Future Sub-GeV Gamma-Ray Telescopes}

\author[a]{Ti-Lin Chou,}\note{ORCID: http://orcid.org/0000-0002-3091-8790}
\emailAdd{tlchou@uchicago.edu}

\author[b]{Dimitrios Tanoglidis}\note{ORCID: http://orcid.org/0000-0002-4631-4529}
\emailAdd{dtanoglidis@uchicago.edu}

\author[b,c,d]{and Dan Hooper}\note{ORCID: http://orcid.org/0000-0001-8837-4127}
\emailAdd{dhooper@fnal.gov}

\affiliation[a]{University of Chicago, Department of Physics, Chicago, IL 60637}
\affiliation[b]{University of Chicago, Department of Astronomy and Astrophysics, Chicago, IL 60637}
\affiliation[c]{Fermi National Accelerator Laboratory, Center for Particle
Astrophysics, Batavia, IL 60510}
\affiliation[d]{University of Chicago, Kavli Institute for Cosmological Physics, Chicago, IL 60637}

\abstract{Annihilating dark matter particles in nearby subhalos could generate potentially observable fluxes of gamma rays, unaccompanied by emission at other wavelengths. Furthermore, this gamma-ray emission is expected to be spatially extended, providing us with a powerful way to discriminate dark matter subhalos from other astrophysical gamma-ray sources. Fermi has detected two dark matter subhalo candidates which exhibit a statistically significant degree of spatial extension (3FGL J2212.5+0703 and 3FGL J1924.8-1034). It has been argued that the most likely non-dark matter interpretation of these observations is that they are each in fact multiple nearby point sources, too close to one another on the sky to be individually resolved. In this study, we consider the ability of next generation gamma-ray telescopes to spatially resolve the gamma-ray emission from subhalo candidates, focusing on the proposed e-ASTROGAM mission. We find that such an instrument could significantly clarify the nature of Fermi's dark matter subhalo candidates, and provide an unprecedented level of sensitivity to the presence of annihilating dark matter in nearby subhalos.}

\maketitle

\section{Introduction}

Indirect searches for dark matter currently employ a wide range of strategies. Among the targets of indirect searches with gamma-ray telescopes are dwarf spheroidal galaxies~\cite{Drlica-Wagner:2015xua,Geringer-Sameth:2014qqa}, the Galactic Center~\cite{Hooper:2012sr}, galaxy clusters~\cite{Lisanti:2017qlb,Ackermann:2015fdi,2010JCAP...05..025A}, and the extragalactic gamma-ray background~\cite{Ackermann:2015tah,DiMauro:2015tfa}. Each of these approaches offers relative advantages and disadvantages.  In particular, the Galactic Center is expected to generate a very bright flux of dark matter annihilation products, while also suffering from significant astrophysical backgrounds. In contrast, the fluxes of dark matter annihilation products from dwarf spheroidal galaxies is predicted to be much lower, but with very low backgrounds.

An alternative strategy is to search for the gamma rays that are produced through dark matter annihilations within one or more nearby subhalos~\cite{Hooper:2016cld,Calore:2016ogv,Bertoni:2015mla,Bertoni:2016hoh,Schoonenberg:2016aml,Berlin:2013dva,Belikov:2011pu,Buckley:2010vg,Ackermann:2012nb,Zechlin:2011wa,Mirabal:2012em,Mirabal:2010ny,Zechlin:2012by,Kuhlen:2008aw,Pieri:2007ir,Baltz:2008wd,Springel:2008by,Springel:2008zz,Koushiappas:2003bn,Tasitsiomi:2002vh}. Throughout this paper, we define a subhalo as a gravitationally bound clump of dark matter located within the halo of the Milky Way. Within the standard paradigm of cold and collisionless dark matter, structures form through repeated mergers, leading the halos that encompass galaxies to contain large numbers of smaller subhalos~\cite{White:1991mr}. Although the largest of the Milky Way's subhalos host dwarf spheroidal galaxies, as well as the Large and Small Magellanic Clouds, a much larger number of smaller subhalos are also expected to be present which do not contain appreciable quantities of gas or stars. 

Annihilating dark matter in nearby subhalos could generate potentially observable fluxes of gamma rays, unaccompanied by radiation at other wavelengths. Estimates for the number of subhalos detectable by the Fermi Gamma-Ray Space Telescope vary considerably, depending on the assumptions that are made about the local distribution of dark matter subhalos and on the shapes of their dark matter profiles. For example, limits placed on the dark matter's annihilation cross section by the authors of Ref.~\cite{Bertoni:2015mla} and Ref.~\cite{Schoonenberg:2016aml} differ by a factor of a few for most dark matter masses, resulting largely from differences in the subhalo density profiles adopted. More specifically, while \Ref{Bertoni:2015mla} adopted density profiles described by a tidally truncated Einasto profile, \Ref{Schoonenberg:2016aml} chose instead to adopt a traditional Navarro-Frenk-White (NFW) density profile, with concentrations selected to match the parameters of a given subhalo identified within the Via Lactea II simulation. A more recent study~\cite{Hooper:2016cld} based on data from the ELVIS and Via Lactea-II simulations favors a result that falls between those presented in Refs.~\cite{Bertoni:2015mla} and~\cite{Schoonenberg:2016aml}, while the study presented in Ref.~\cite{Calore:2016ogv} predicts a somewhat smaller number of detectable subhalos. Regardless of which of these results is closer to the correct answer, it is generally agreed that if dark matter consists of particles with a mass in the range of $\sim$10-100 GeV and with an annihilation cross section similar to that naively predicted for a thermal relic ($\sigma v \simeq 2 \times 10^{-26}$ cm$^3/$s), a small number of nearby subhalos could plausibly be detectable by Fermi, as well as by future space-based gamma-ray telescopes.

The Fermi Collaboration's Third Source Catalog (3FGL) contains 992 gamma-ray sources that have not been associated with emission observed at other wavelengths~\cite{TheFermi-LAT:2015hja}, including 19 high-latitude  ($|b|>20^{\circ}$) sources that are bright ($\Phi_{\gamma}>7 \times 10^{-10}$ cm$^{-2}$ s$^{-1}$, $E_{\gamma} > 1$ GeV), non-variable, and that exhibit a spectral shape that is consistent with the predictions of annihilating dark matter~\cite{Bertoni:2015mla,Bertoni:2016hoh}. Although it is likely that most of these objects are unidentified astrophysical sources, such as radio-faint pulsars, it is plausible that some of them could be dark matter subhalos. 

A powerful way to potentially discriminate a dark matter subhalo from an unidentified astrophysical source is to study the morphology of the associated gamma-ray emission. In particular, it has been argued that a robust detection of a spatially extended gamma-ray source without observable counterparts at other wavelengths would constitute a smoking gun for annihilating dark matter~\cite{Bertoni:2016hoh}. While diffuse emission mechanisms (pion production, inverse Compton scattering, and bremsstrahlung) can generate spatially extended gamma-ray emission from astrophysical sources, these processes also invariably generate bright emission at other wavelengths, through processes such as synchrotron, or through the heating of diffuse material.  Furthermore, in contrast to more compact objects, any multi-wavelength emission that is generated in diffuse environments will not be readily absorbed or significantly beamed. It is noteworthy that very bright multi-wavelength emission has been detected~\cite{Green:2014cea,TheFermi-LAT:2015hja,chandracatalog,Dubner:2013bpa,Kargaltsev:2007kf,2013ApJ...771...91B,2012ApJ...752..135K,2013MNRAS.436..968L,2001AAS...198.4002L,Kargaltsev:2008hf,Sakai:2011sv,Acero:2013xta,Nobukawa:2015jva,Mizuno:2015jua} from all 25 of the sources in the 3FGL catalog~\cite{TheFermi-LAT:2015hja} that are classified as spatially extended. The majority of these extended sources are supernova remnants and pulsar wind nebula, along with the star-forming region Cygnus X, the lobes of the radio galaxy Centaurus A, and the Large and Small Magellanic Clouds. 

It was recently pointed out in Refs.~\cite{Bertoni:2015mla,Bertoni:2016hoh} (see also Ref.~\cite{Wang:2016xjx}) that the gamma-ray emission from the unassociated Fermi source 3FGL J2212.5+0703 is spatially extended, with a radius of $\sim$0.2$^{\circ}$, preferring an extended profile over that of a point source with a significance of 5.1$\sigma$. Furthermore, this source is located far from the Galactic Plane ($b=-38.56$), shows no signs of variability, and exhibits a spectral shape that is well fit by the annihilations of an $\sim$15-35 GeV dark matter candidate (for the case of annihilations to $b\bar{b}$). Even more recently, the source 3FGL J1924.8-1034 was also shown at the 5.4$\sigma$ confidence level to be extended, with a best fit radius of $\sim$0.15$^{\circ}$, and favoring a similar range of dark matter masses as 3FGL J2212.5+0703~\cite{Xia:2016uog}.

The lack of observable multi-wavelength counterparts make it very unlikely that either 3FGL J2212.5+0703 or 3FGL J1924.8-1034 are extended astrophysical sources. Instead, the most likely non-dark matter interpretation of these observations is that these apparently extended sources are each, in fact, multiple point sources, too close to one another on the sky to be individually resolved by Fermi~\cite{Bertoni:2016hoh,Xia:2016uog}. In Ref.~\cite{Bertoni:2016hoh}, it was estimated that there is a small but non-negligible chance probability (approximately $\sim$2\%) that a pair of such sources would appear somewhere on the sky. In light of this, it is difficult at this time to make a convincing case that either 3FGL J2212.5+0703 or 3FGL J1924.8-1034 is a dark matter subhalo. 

In order to definitively resolve the spatial morphology of the gamma-ray emission associated with these apparently extended sources, it is likely that instruments beyond Fermi will be required. There are plans, however, for future space-based gamma-ray telescopes which would be very useful in this regard. Such proposals include the All-sky Medium Energy Gamma-ray Observatory (AMEGO), the Compton-Pair Production Space Telescope (ComPair)~\cite{Moiseev:2015lva}, and enhanced ASTROGAM (e-ASTROGAM)~\cite{DeAngelis:2016slk,Tatischeff:2016ykb}.\footnote{{\url https://asd.gsfc.nasa.gov/amego/index.html}, \, \, \, {\url http://www.iaps.inaf.it/eastrogam/}} These instruments are each designed to utilize both Compton scattering and pair production signals, offering unprecedented sensitivity to photons in the MeV-GeV energy range. 

\begin{figure}[t]
\centering
\includegraphics[scale=0.431]{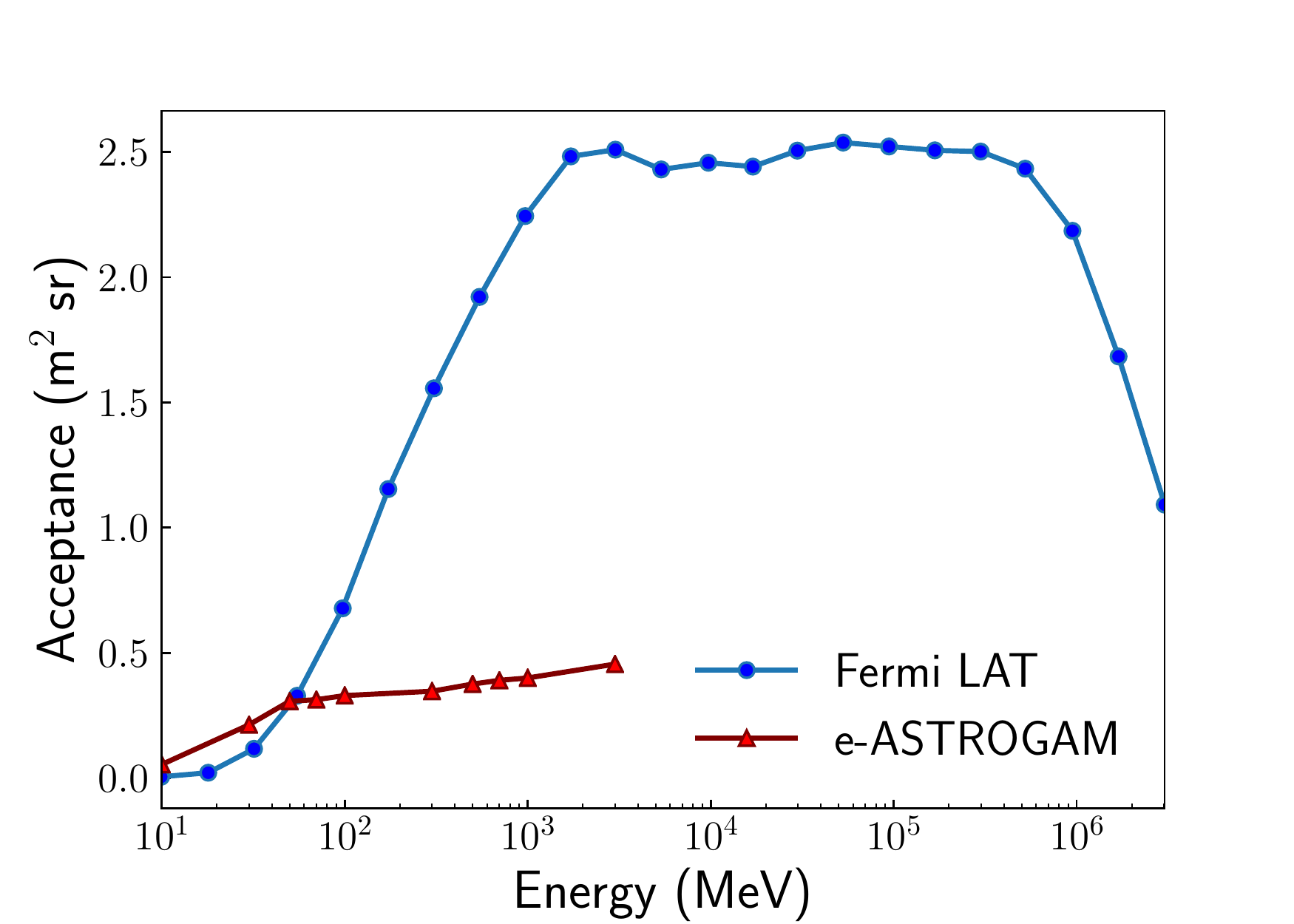}
\includegraphics[scale=0.431]{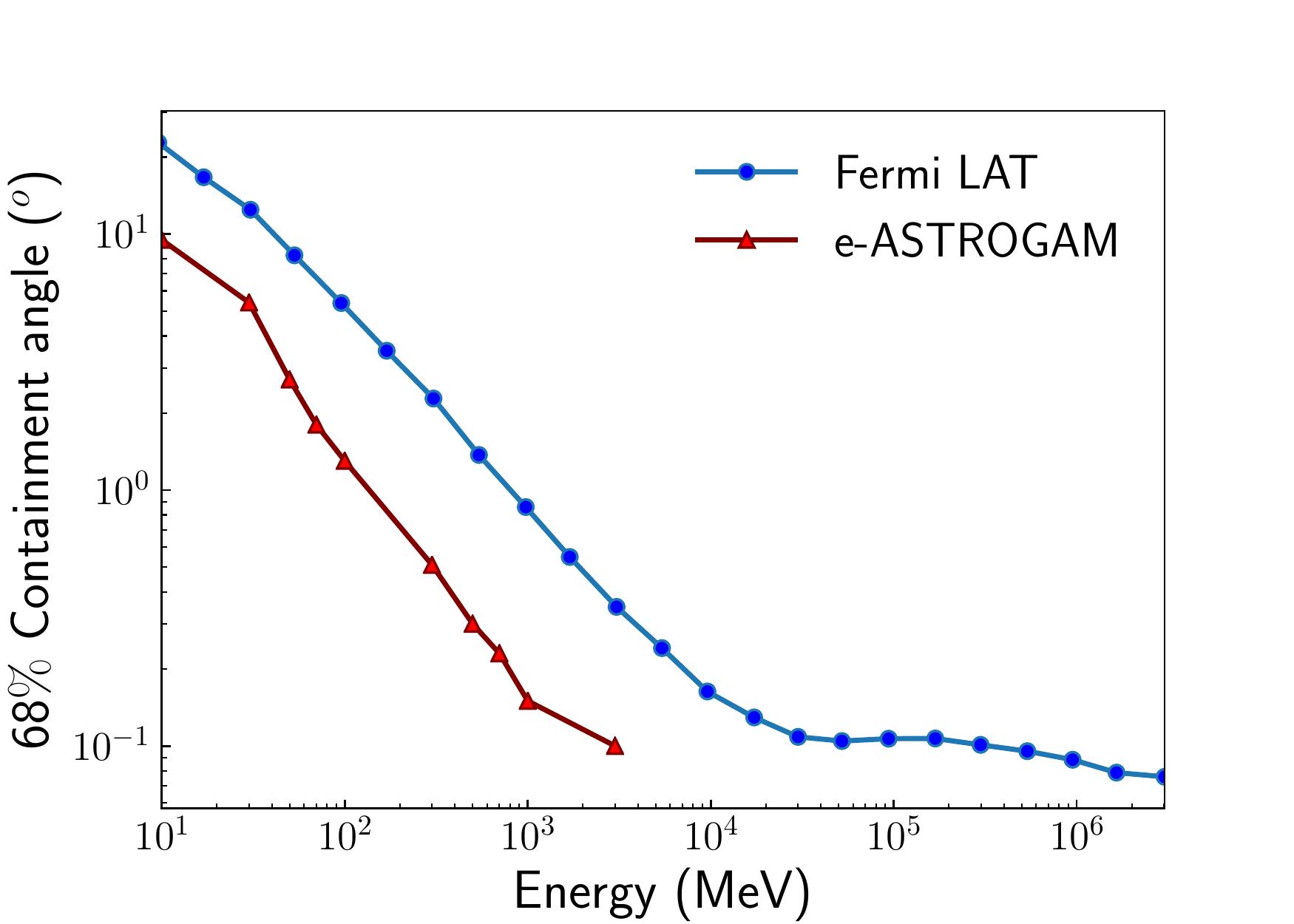}
\caption{The projected accceptance (left) and 68\% containment radius (right) of e-ASTROGAM, compared to that of the Fermi's Large Area Telescope (LAT). Although the total acceptance of e-ASTROGAM is only competitive with that of Fermi at energies below $\sim$100 MeV, the improvement in angular resolution (by a factor of $\sim$4-6
at energies between 100 MeV and 1 GeV) will be invaluable in ascertaining the nature of Fermi's dark matter subhalo candidates.}
\label{acceptance}
\end{figure}

For concreteness, we focus in this study on the case of e-ASTROGAM. In Fig.~\ref{acceptance} we present the projected acceptance and 68\% angular containment radius for e-ASTROGAM~\cite{DeAngelis:2016slk}, compared to that for the Fermi Gamma-Ray Space Telescope.\footnote{{\url http://www.slac.stanford.edu/exp/glast/groups/canda/lat$\_$Performance.htm}} In the case of e-ASTROGAM, we took the acceptance to be equal to the effective area reported in Ref.~\cite{DeAngelis:2016slk} multiplied by $0.2 \times 4\pi$ steridians, representing the telescope's approximate field-of-view. Although the total acceptance of e-ASTROGAM is competitive with that of Fermi only at energies below $\sim$100 MeV, the superior angular resolution of this experiment would be invaluable in ascertaining the nature of the subhalo candidate sources 3FGL J2212.5+0703 and 3FGL J1924.8-1034.

Previous studies have considered the prospects for detecting annihilating or decaying dark matter with an experiment such as e-ASTROGAM, focusing on the case of relatively light (\ie sub-GeV) dark matter candidates, and their novel spectral features~\cite{Bartels:2017dpb,Bringmann:2016axu,Pittori:2016pre}. Here, we instead focus on e-ASTROGAM's ability to detect and resolve the morphology of the dark matter subhalo candidates previously observed by Fermi.

\section{Dark Matter Subhalos and Future Sub-GeV Gamma-Ray Telescopes}

\begin{figure}[h]
\centering
\includegraphics[scale=0.55]{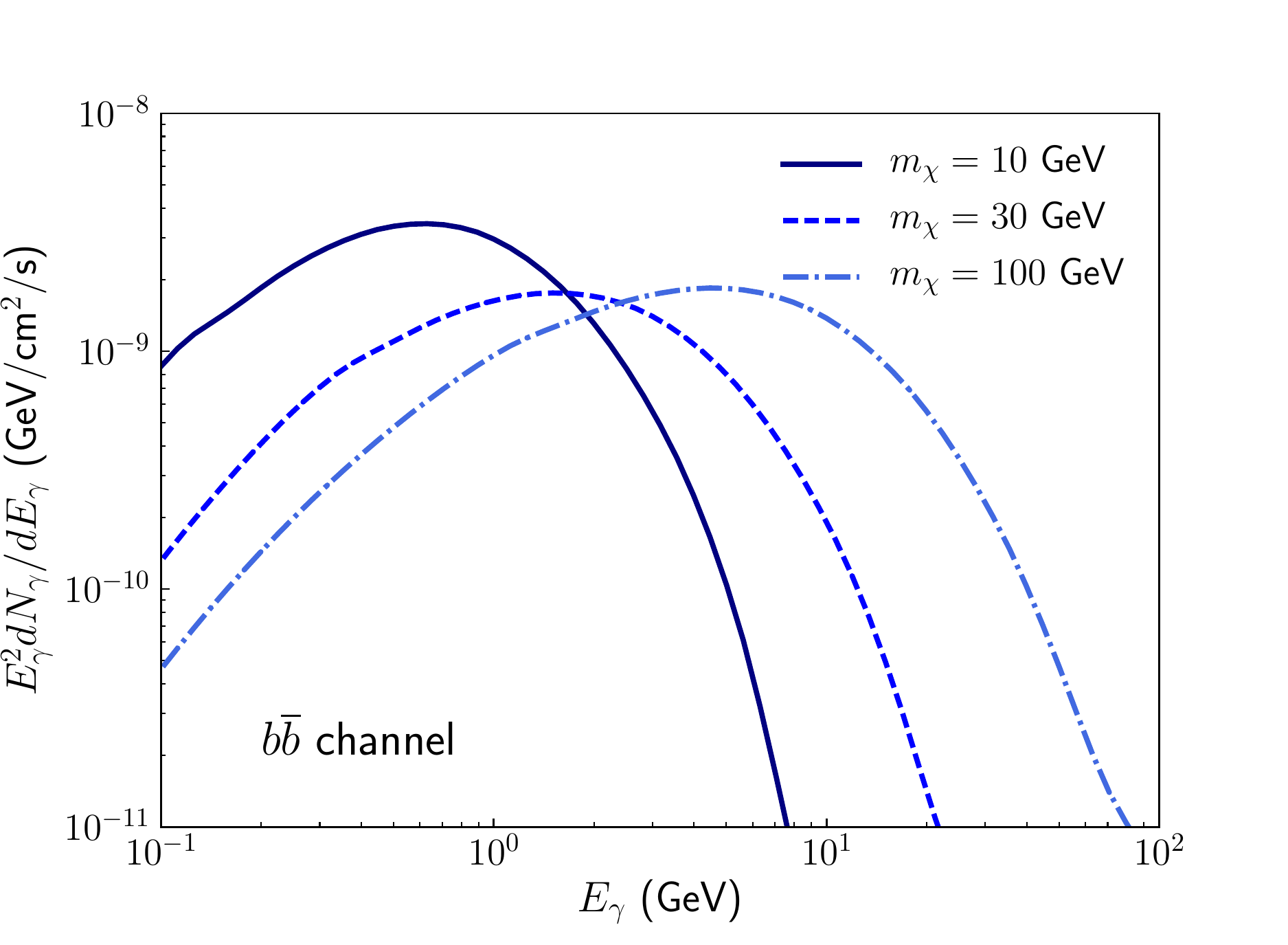}
\caption{The gamma-ray spectrum from dark matter annihilating to $b\bar{b}$, for three choices of the mass. In each case, the overall flux is normalized to $1.33\times 10^{-9}$ cm$^{-2}$ s$^{-1}$ above 1 GeV (approximately equal to the flux observed from 3FGL J2212.5+0703~\cite{Bertoni:2016hoh}).}
\label{spectrum}
\end{figure}

The gamma-ray emission from a given dark matter subhalo is described as follows:
\begin{equation}
\Phi_{\gamma}(E_{\gamma},\theta) = \frac{\langle \sigma v \rangle}{8 \pi m^2_\chi} \frac{dN_{\gamma}}{dE_{\gamma}} \int_{\rm los} \rho^2(r) \, dl,
\end{equation}
where $\left<\sigma v \right>$ is the dark matter's thermally averaged annihilation cross section, $m_\chi$ is the dark matter's mass, $dN_\gamma/dE_{\gamma}$ is the spectrum of gamma rays produced per annihilation, and the integral is carried out over the line-of-sight (los). The radial density profile, $\rho(r)$, is given in terms of the distance to the center of the subhalo, $r=\sqrt{l^2+D^2-2lD\cos \theta}$, where $D$ is the distance from the subhalo to the Solar System. Following Ref.~\cite{Bertoni:2016hoh}, we adopt a subhalo density profile that is described by an NFW distribution, tidally truncated at a radius such that only the innermost 0.5\% of the mass remains intact~\cite{Springel:2008cc}. Throughout this study, we quantify the degree of spatial extension from a given gamma-ray source by the quantity, $\sigma_{_{68}}$, which is defined as the angular radius that contains $68\%$ of the total photons from that source. For a given dark matter mass and annihilation channel, we calculate $dN_{\gamma}/dE_{\gamma}$ using PYTHIA 8~\cite{pythia8} (see also Ref.~\cite{cirelli2011pppc}). In Fig.~\ref{spectrum} we show the gamma-ray spectrum produced from dark matter annihilating to $b\bar{b}$, for three values of the mass, and in each case normalized to a flux of $1.33\times 10^{-9} \, {\rm cm}^{-2} \, {\rm s}^{-1}$ above 1 GeV (the approximate flux observed from 3FGL J2212.5+0703).

\begin{figure}[t]
\centering
\includegraphics[scale=0.55]{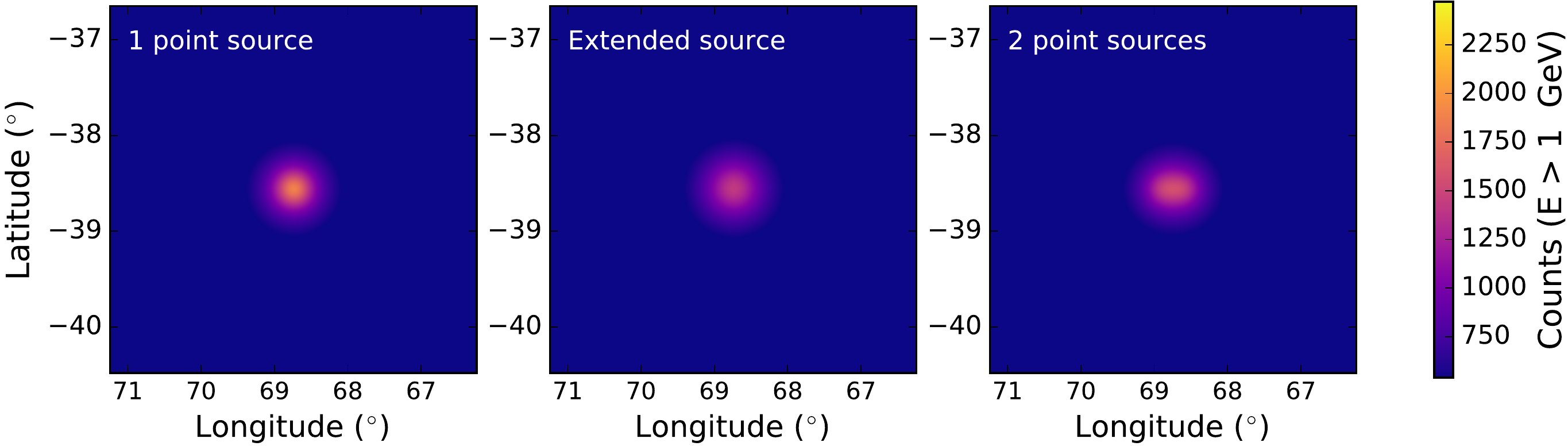} \\
\vspace{0.5cm}
\includegraphics[scale=0.55]{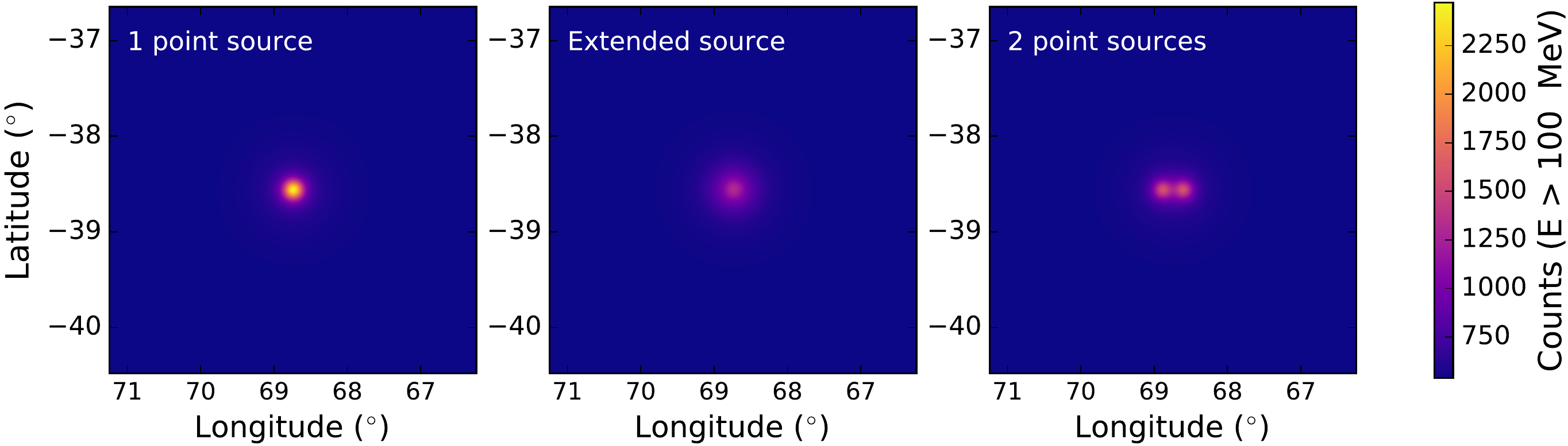}
\caption{Simulated photon-count maps as measured by either Fermi LAT (top) or e-ASTROGAM (bottom). In the left frames, the maps correspond to the emission from a single point source, while the center and right frames depict images from a subhalo-like extended source (with $\sigma_{_{68}}=0.25^{\circ}$) and from a pair of point sources (separated by $0.28^{\circ}$), respectively. In each case, the spectral shape and total flux is equal to that shown in Fig.~\ref{spectrum} for the case of $m_{\chi}=30$ GeV.}
\label{maps}
\end{figure}

In order to evaluate the ability of e-ASTROGAM and/or Fermi to distinguish one or more gamma-ray point sources from an extended source, we have simulated sources of various morphology and spectrum. In each case, we create a $10^\circ\times10^\circ$ map of the emission, divided into $0.0125^{\circ} \times 0.0125^{\circ}$ angular bins, and 20 flux bins evenly spaced in logarithm between 0.01 and 100 GeV. In addition to the sources themselves, we include a background of Galactic diffuse emission~\cite{ackermann2012fermi} and the extragalactic gamma-ray background~\cite{ackermann2015spectrum}. We adopt the instrumental acceptances as shown in the left frame of Fig.~\ref{acceptance}, and consider observations over a period of 10 and 5 years for Fermi and e-ASTROGAM, respectively. The underlying angular distribution of the photons is then convolved with a 2D Gaussian point spread function, with a containment radius for each instrument as shown in the right frame of Fig.~\ref{acceptance}.

In the leftmost frames of Fig.~\ref{maps}, we show simulated photon-count maps for a single point source with a spectral shape and flux equal to that shown in Fig.~\ref{spectrum} for the case of a 30 GeV dark matter particle, as measured by either Fermi LAT (top) or e-ASTROGAM (bottom). In the center frames of this figure, we show simulated maps for an extended gamma-ray source, with a morphology of a dark matter subhalo with a tidally truncated NFW profile, extended to a degree corresponding to $\sigma_{_{68}}=0.25^{\circ}$. Lastly, in the right frames, we show simulated maps for a pair of nearby gamma-ray point sources, of equal flux and separated from one another by a distance of $0.28^{\circ}$.\footnote{Here and throughout this study, we consider point source pairs that are separated by an angle that is chosen to be maximally difficult to distinguish from the case of a single extended source. In the case of an extended subhalo with $\sigma_{_{68}}=0.25^{\circ}$, this corresponds to a separation of $0.28^{\circ}$, whereas for $\sigma_{_{68}}=0.10^{\circ}$ $(0.05^{\circ})$, we find the maximally indistinguishable separation to be $0.12^{\circ}$ ($0.06^{\circ}$).} Whereas it is rather difficult to distinguish the extended source morphology from that of two nearby point sources in the simulated Fermi maps, the difference is much more clear in the simulated e-ASTROGAM images.

To access the ability of Fermi and/or e-ASTROGAM to distinguish between these different morphologies, we use a given map and draw from a Poisson distribution for each bin to produce a series of mock observations of the region. We then calculate the mean log-likelihood with which these mock observations are described by a given model. The log-likelihood is given as follows:
\begin{equation*}
\ln{\mathcal{L}}=\Sigma_i (k_i \ln{\lambda_i}-\lambda_i-\ln{k_i!}) \, ,
\end{equation*}
where the sum is carried out over all angular and energy bins, $k_i$ denotes the number of events in bin $i$, and $\lambda_i$ is the number of events predicted by the model in the same bin. We then define the test statistic (TS) that one model can be distinguished from another as twice the difference in the log-likelihood.

For the case shown in Fig.~\ref{maps}, we find that Fermi alone can distinguish between the single point source model and the extended model at a level of TS=22.3 (similar to the value of TS=21.4 that was found for 3FGL J2212.5+0703 in Ref.~\cite{Bertoni:2016hoh}). But Fermi has a more difficult time distinguishing an extended source from a pair of nearby point sources (TS=11.5, and requiring additional degrees-of-freedoms relative to the single extended source model). In this respect, data from an experiment such as e-ASTROGAM could be very clarifying. In particular, we find that the combination of Fermi and e-ASTROGAM can identify extension from such a source at a level of TS=79.5, and differentiate two point sources from a single extended source with TS=53.5. This represents a clear and qualitative improvement to that possible using Fermi alone.

\begin{figure}[t]
\centering
\includegraphics[scale=0.45]{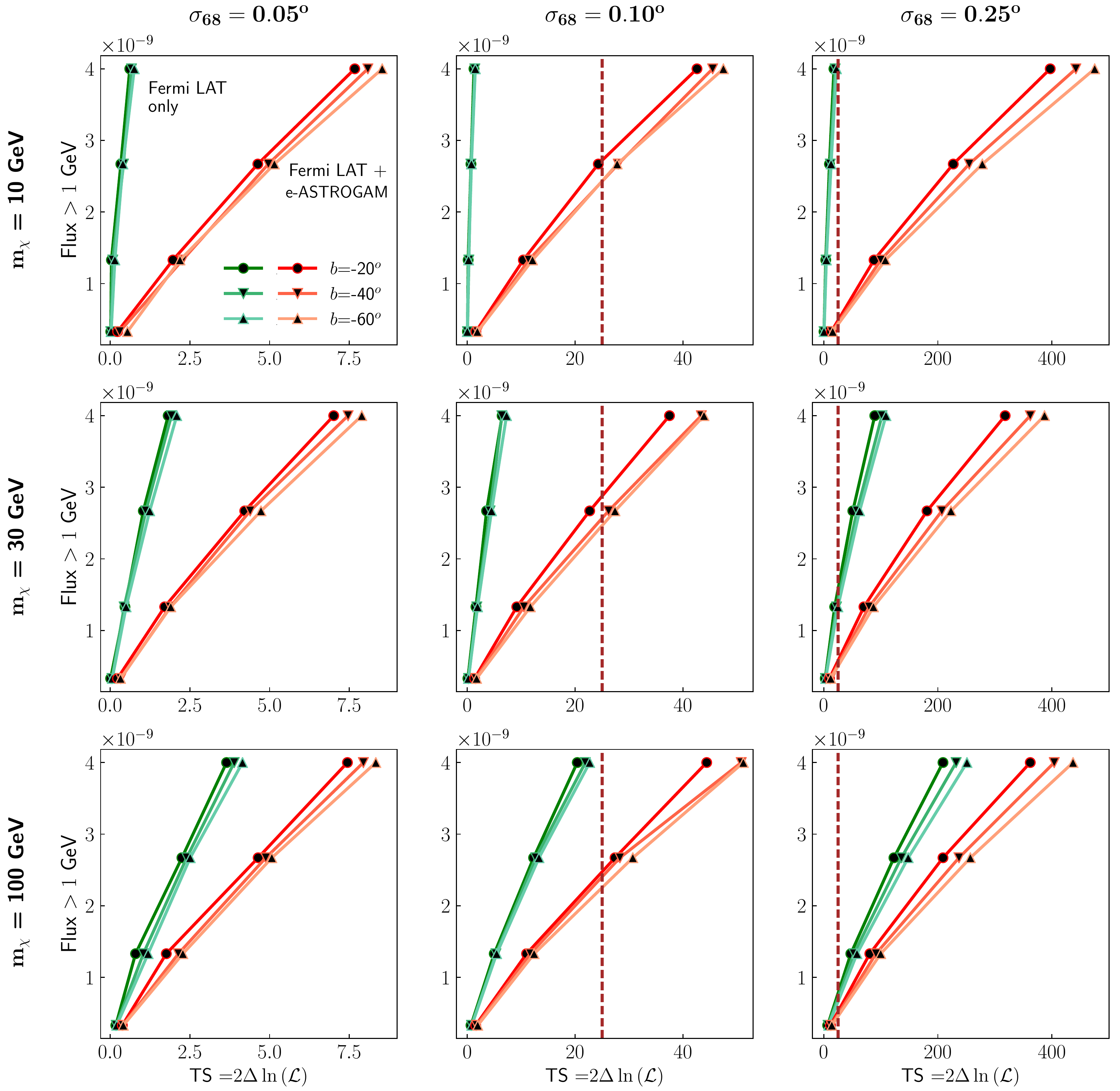}
\caption{The ability of Fermi-LAT alone, and Fermi in conjunction with e-ASTROGAM, to distinguish between a gamma-ray point source and a dark matter subhalo extended at a level of $\sigma_{_{68}}=0.05^{\circ}$, $0.1^{\circ}$ or $0.25^{\circ}$. Results are shown as a function of the flux of the source (in units of photons cm$^{-2}$ s$^{-1}$), for three values of the Galactic latitude, and for the three choices of the dark matter mass shown in Fig.~\ref{spectrum}. The red vertical dashed lines denote TS=25.}
\label{NFW-1PS}
\end{figure}

\begin{figure}[t]
\centering
\includegraphics[scale=0.45]{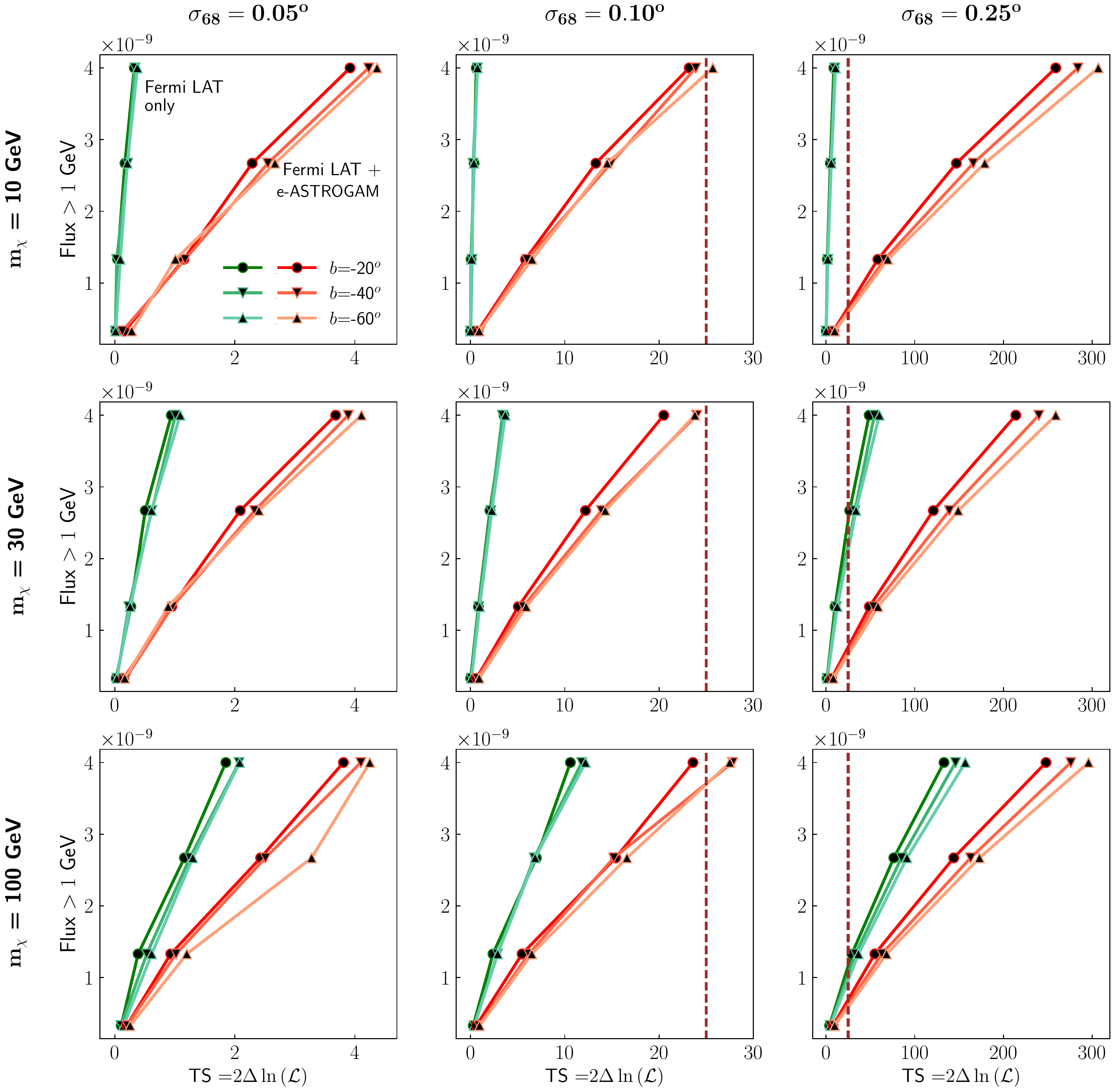}
\caption{As in Fig.~\ref{NFW-1PS}, but assessing the ability of Fermi and e-ASTROGAM to distinguish a single extended dark matter subhalo from a pair of nearby gamma-ray point sources.}
\label{NFW-2PS}
\end{figure}

In Figs.~\ref{NFW-1PS} and~\ref{NFW-2PS}, we generalize these results, plotting the ability of either Fermi alone, or Fermi combined with e-ASTROGAM, to differentiate a gamma-ray point source, or a pair of point sources, from an extended source. We show results for subhalos extended with $\sigma_{_{68}}=0.05^{\circ}$, $0.1^{\circ}$ or $0.25^{\circ}$, and for spectral shapes corresponding to the three dark matter models shown in Fig.~\ref{spectrum}. For each dark matter model and extension, we plot the ability of these instruments to distinguish the morphologies (presented in terms of the test statistic, TS) as a function of the total flux of the gamma-ray source(s). We also show these results for three different choices of the Galactic Latitude (the diffuse gamma-ray background is lower at higher latitudes).

We would like to emphasis two important points that are illustrated by Figs.~\ref{NFW-1PS} and~\ref{NFW-2PS}. First, although Fermi may sometimes be unable to distinguish an extended gamma-ray source from a pair of nearby sources, e-ASTROGAM will in many cases be capable of breaking such degeneracies (as shown in Fig.~\ref{NFW-2PS}). Such an instrument would be expected to shed considerable light on the sources 3FGL J2212.5+0703 or 3FGL J1924.8-1034, resolving their morphologies in some detail. Second, if either or both of the extended gamma-ray sources 3FGL J2212.5+0703 or 3FGL J1924.8-1034 are in fact dark matter subhalos, we should expect e-ASTROGAM to identify spatially extended gamma-ray emission from several more subhalo candidates. For example, in the case of a dark matter mass of 30 GeV and annihilations to $b\bar{b}$, Fig.~\ref{NFW-1PS} indicates that Fermi can identify the presence of extension from a subhalo (with TS $\gsim$ 25) only for very bright sources ($F_{\gamma} \gsim 10^{-9}$ cm$^{-2}$ s$^{-1}$ above 1 GeV) with $\sigma_{_{68}} \gsim 0.2-0.3$. But the combination of Fermi and e-ASTROGAM would be able to identify the extension of a subhalo with a similar luminosity and physical extent that is located at a distance of up to $\sim$30-40\% farther away. Thus for a locally homogeneous distribution of subhalos, we expect e-ASTROGAM to increase the number of subhalos with detectable spatial extension by a factor of $\sim$2-3. Within this context, particularly promising targets for e-ASTROGAM include the Fermi sources 3FGL J1119.9-2204 and 3FGL J0381.1+0252, each of which currently exhibit hints of spatial extension (TS $\simeq$ 7.7 and 5.8, respectively)~\cite{Bertoni:2016hoh}. If these sources are in fact extended at a level of $\sigma_{_{68}}\sim0.07^{\circ}$, $0.15^{\circ}$ as suggested by the Fermi data, this extension would be expected to be resolved with high significance by e-ASTROGAM.

\section{Summary and Conclusions \label{sec:conc}}

In this study, we have considered the prospects for e-ASTROGAM to improve upon our ability to characterize the morphology of spatially extended gamma-ray sources. Although Fermi has detected a small number of sources which appear to be spatially extended with a radius of $\sim$0.2-0.3$^{\circ}$, it is possible that these observations are, in fact, of multiple point sources which are too close to each other to be resolved. Due to e-ASTROGAM's high angular resolution at $\sim$0.1-1 GeV energies, it is expected that such an instrument would be able to clearly resolve such sources, and considerably clarify their nature. Furthermore, if any of Fermi's extended sources are in fact dark matter subhalos, we expect e-ASTROGAM to identify spatially extended gamma-ray emission from several more subhalo candidates.

\bigskip
\bigskip

\textbf{Acknowledgments.} This work was performed in part at the Aspen Center for Physics, which is supported by National Science Foundation grant PHY-1607611. This manuscript has been authored in part by the Fermi Research Alliance, LLC under Contract No. DE-AC02-07CH11359 with the U.S. Department of Energy, Office of Science, Office of High Energy Physics. The United States Government retains and the publisher, by accepting the article for publication, acknowledges that the United States Government retains a non-exclusive, paid-up, irrevocable, world-wide license to publish or reproduce the published form of this manuscript, or allow others to do so, for United States Government purposes.

\bibliographystyle{JHEP}
\bibliography{astrogam}

\end{document}